\documentclass[prb,reprint,amsfonts,superscriptaddress]{revtex4-1}

\usepackage{amsmath}
\usepackage{amssymb}
\usepackage{amsfonts}
\usepackage[pdftex]{graphicx}
\usepackage{units}
\usepackage[usenames]{color}
\usepackage{dcolumn}
\usepackage{bm}
\usepackage{float}
\usepackage{xspace}
\usepackage[usenames]{color}
\usepackage{textcomp}
\usepackage[]{placeins}

\def\Ef{$E_{\rm F}$}
\def\Eb{$E_{\rm B}$}
\def\Ed{$E_{\rm D}$}

\def\kf{$k_{\rm F}$}
\def\kz{$k_{\rm z}$}

\def\kpara{{\bf k}$_\parallel$}
\def\kperp{{\bf k}$_\perp$}

\def\kperp{{\bf k}$_\perp$}
\def\invA{\AA$^{-1}$}

\def\Xbar{$\overline{\rm X}$}
\def\Ybar{$\overline{\rm Y}$}

\def\GbarMbar{$\overline{\Gamma}$-$\overline{\rm M}$}
\def\GbarXbar{$\overline{\Gamma}$-$\overline{\rm X}$}

\def\etal{\textit{et al.}}

\def\BiTe{Sb$_2$Te$_3$}
\def\BiTe{Bi$_2$Te$_3$}
\def\BiSe{Bi$_2$Se$_3$}
\def\BiX{Bi$_2$X$_3$}
\def\aSn{$\alpha$-Sn}

\def \Z{$\mathbb{Z}_2$}

\newcommand{\GSM}{$ \mathrm{\Gamma_{7}^{-}} $\xspace} 
\newcommand{\GSP}{$ \mathrm{\Gamma_{7}^{+}} $\xspace} 
\newcommand{\GEP}{$ \mathrm{\Gamma_{8}^{+}} $\xspace} 

\newcommand{\IMS}{Im$\Sigma$ \xspace}

\begin{document}
\title{The topological surface state of \aSn\ on InSb(001) as studied by  photoemission}

\author{M. R. Scholz}
\affiliation{\mbox{Physikalisches Institut and R\"ontgen Center for Complex Material Systems, Universit\"at W\"urzburg, 97074 W\"urzburg, Germany}}

\author{V. A. Rogalev}
\affiliation{\mbox{Physikalisches Institut and R\"ontgen Center for Complex Material Systems, Universit\"at W\"urzburg, 97074 W\"urzburg, Germany}}

\author{L. Dudy}
\affiliation{\mbox{Physikalisches Institut and R\"ontgen Center for Complex Material Systems, Universit\"at W\"urzburg, 97074 W\"urzburg, Germany}}

\author{F. Reis}
\affiliation{\mbox{Physikalisches Institut and R\"ontgen Center for Complex Material Systems, Universit\"at W\"urzburg, 97074 W\"urzburg, Germany}}

\author{F. Adler}
\affiliation{\mbox{Physikalisches Institut and R\"ontgen Center for Complex Material Systems, Universit\"at W\"urzburg, 97074 W\"urzburg, Germany}}

\author{\mbox{J. Aulbach}}
\affiliation{\mbox{Physikalisches Institut and R\"ontgen Center for Complex Material Systems, Universit\"at W\"urzburg, 97074 W\"urzburg, Germany}}


\author{L. J. Collins-McIntyre}
\affiliation{Clarendon Laboratory, Physics Department, Oxford University, OX1 3PU, United Kingdom}

\author{L. B. Duffy}
\affiliation{Clarendon Laboratory, Physics Department, Oxford University, OX1 3PU, United Kingdom}

\author{H. F. Yang}
\affiliation{Clarendon Laboratory, Physics Department, Oxford University, OX1 3PU, United Kingdom}
\affiliation{School of Physical Science and Technology, ShanghaiTech University, Shanghai 201210, China}

\author{Y. L. Chen}
\affiliation{Clarendon Laboratory, Physics Department, Oxford University, OX1 3PU, United Kingdom}

\author{T. Hesjedal}
\affiliation{Clarendon Laboratory, Physics Department, Oxford University, OX1 3PU, United Kingdom}

\author{Z. K. Liu}
\affiliation{Diamond Light Source, Didcot, OX11 0DE, United Kingdom}
\affiliation{School of Physical Science and Technology, ShanghaiTech University, Shanghai 201210, China}

\author{M. Hoesch}
\affiliation{Diamond Light Source, Didcot, OX11 0DE, United Kingdom} %

\author{S. Muff}
\affiliation{Swiss Light Source, Paul Scherrer Institut, CH-5232 Villigen, Switzerland}
\affiliation{Institute of Physics, Ecole Polytechnique F\'{e}d\'{e}rale de Lausanne, CH-1015 Lausanne, Switzerland}

\author{J. H. Dil}
\affiliation{Swiss Light Source, Paul Scherrer Institut, CH-5232 Villigen, Switzerland}
\affiliation{Institute of Physics, Ecole Polytechnique F\'{e}d\'{e}rale de Lausanne, CH-1015 Lausanne, Switzerland}

\author{J. Sch{\"a}fer}
\affiliation{\mbox{Physikalisches Institut and R\"ontgen Center for Complex Material Systems, Universit\"at W\"urzburg, 97074 W\"urzburg, Germany}}

\author{R. Claessen}
\affiliation{\mbox{Physikalisches Institut and R\"ontgen Center for Complex Material Systems, Universit\"at W\"urzburg, 97074 W\"urzburg, Germany}}

\begin{abstract}
We report on the electronic structure of the elemental topological semimetal \aSn\ on InSb(001). High-resolution angle-resolved photoemission data allow to observe the topological surface state (TSS) that is degenerate with the bulk band structure and show that the former is unaffected by different surface reconstructions. An unintentional $p$-type doping of the as-grown films was compensated by deposition of potassium or tellurium after the growth, thereby shifting the Dirac point of the surface state below the Fermi level. We show that, while having the potential to break time-reversal symmetry, iron impurities with a coverage of up to 0.25 monolayers do not have any further impact on the surface state beyond that of K or Te. Furthermore, we have measured the spin-momentum locking of electrons from the TSS by means of spin-resolved photoemission. Our results show that the spin vector lies fully in-plane, but it also has a finite radial component. Finally, we analyze the decay of photoholes introduced in the photoemission process, and by this gain insight into the many-body interactions in the system. 
Surprisingly, we extract quasiparticle lifetimes comparable to other topological materials where the TSS is located within a bulk band gap. We argue that the main decay of photoholes is caused by intraband scattering, while scattering into bulk states is suppressed due to different orbital symmetries of bulk and surface states.
\end{abstract}
\maketitle

\section{\label{Intro} Introduction}

The low-temperature $\alpha$-phase of Sn has attracted considerable attention recently as a unique elemental three-dimensional topologically non-trivial material \cite{,Fu:2007ei,Barfuss:2013by,Ohtsubo:2013je,RojasSanchez:2016gr,Xu2017,Rogalev:2016tz,Huang2017}. Being a zero-gap semiconductor if unstrained, it can enter a Dirac semimetal or strong topological insulator (TI) phases under strain \cite{Rogalev:2016tz,Huang2017}. Interestingly, the non-trivial topology and, hence, the topological surface state (TSS) in \aSn\ exist independently of the strain owing to a robust band inversion between conduction and second valence bands, similar to HgTe \cite{Bernevig2006} or some ternary Heusler compounds \cite{Chadov2010}. While the electronic structure of \aSn\ has been reported in several experimental studies \cite{Barfuss:2013by,Ohtsubo:2013je,Xu2017,RojasSanchez:2016gr,Rogalev:2016tz}, a detailed analysis of the TSS is still missing. 

Despite the degeneracy of the TSS with the projected bulk bands, \aSn\ has proven its potential for spintronics applications \cite{RojasSanchez:2016gr}. A weak hybridization of the TSS with the energetically coexisting bulk bands \cite{Rogalev:2016tz} calls for the analysis of the TSS quasiparticle lifetime, which has not been reported so far.
Additional spintronic functionalities can be expected from introducing ferromagnetic impurities, which are  suggested to break time-reversal symmetry, thereby opening a gap in the TSS. This is of interest for devices, such as transistors, as well as for achieving the quantum anomalous Hall state \cite{Qi:2008ub,Liu:2009ig,Gao:2009da,Yu:2010fr,Mondal:2010ju,Garate:2010fh,Mondal:2010kx,Kong:2011ks}.
However, their experimental realization has proven challenging and has led to yet another controversy in the literature, i.e., whether adding ferromagnetic impurities in a TI can indeed open a band gap in the TSS \cite{Chen:2010jv,Wray:2010te,Bianchi:2011kc,Ye:2012df,Scholz:2012kz,Honolka:2012kp,Xu:2012fj,Scholz:2013gl,Sessi2016dual}.
Such effects were mainly studied in the \BiX\ family of compounds, and related ternary systems, and it has been argued that, e.g., the high intrinsic doping in \BiX\ prevents the proposed surface-state-mediated ferromagnetic alignment of the impurities \cite{Liu:2009ig,Scholz:2012kz}. Hence, the tunability of the Dirac point might be an essential requirement to obtain and study a band gap opening via magnetic impurities in the TSS.

In this paper, we present different aspects of the electronic structure of \aSn\ films using spin- and angle-resolved photoemission (ARPES). We reveal that although the TSS is degenerate with the bulk band structure, the former maintains an isotropic \kpara-space dispersion with circular constant energy contours (CECs) unaffected by different surface reconstructions of the film.
We compensate the intrinsic $p$-type doping of the as-grown film by surface deposition of potassium, thereby shifting the Kramers degeneracy point of the surface state below the Fermi level. Additional control over the precise position of \Ef\ is gained through the variation of the thickness of the Te buffer layer between substrate and film.
Despite this tunability, we present data that unambiguously demonstrates the absence of a resolvable gap in the TSS in \aSn\ after Fe adatom deposition [up to 0.25 monolayers (ML)], which is in line with previous findings in \BiSe\ and \BiTe\ \cite{Scholz:2012kz,Scholz:2013gl}.
Furthermore, we have measured the spin-vector alignment of electrons in the TSS by means of spin-resolved ARPES. Our results show that the spin vector lies fully in the sample plane along all directions in $k$-space.
Further, our results on the circular dichroism in the angular distribution (CDAD) of photoelectrons confirm that the method does not provide a reliable measure of spin polarization \cite{Scholz:2013fk,SanchezBarriga:2014fj}.
Finally, we analyze the decay of photoholes introduced in the photoemission process, and in this way, get insight into the many-body interactions in the system. The extracted quasiparticle lifetimes are found to be comparable to other TIs where the TSSs appear within a bulk band gap. We argue that the main decay of photoholes is caused by intraband scattering and that scattering into bulk states is suppressed by virtue of different orbital symmetries and consequently small hybridization of bulk and surface states.

\section{\label{Methods} Methods}
\aSn\ thin film samples were grown by molecular beam epitaxy (MBE) on InSb(001) substrates. The 8-effusion cell MBE system is directly attached to the high-resolution ARPES system at beamline I05 at the Diamond Light Source (Didcot, UK), allowing for in-vacuum transfers \cite{Baker:2015}.
The substrates were cleaned by several sputter and anneal cycles until a clear c(8$\times$2)-reconstruction was observed by low-energy-electron diffraction (LEED). Afterwards, a Te buffer layer was deposited, inducing a (1$\times$1) surface reconstruction.
The amount of Te was monitored by x-ray photoemission spectroscopy (XPS) and LEED.
Next, Sn was grown layer-by-layer, as evidenced by clear RHEED oscillations (not shown here). 
After depositing for $\sim$10-12 RHEED oscillations (monatomic Sn layers) the thickness of the film is sufficient to neglect the influence of the InSb/Te interface.

ARPES measurements have been carried out with $s$ and $p$ linearly polarized light, as well as circularly polarized light, at varying photon energies at beamline I05. The endstation is equipped with a Scienta R4000 hemispherical electron analyzer that provides an ultimate energy and angular resolution of $\sim$5\,meV and 0.1$^{\circ}$, respectively. Spin-resolved photoemission spectra were measured at the SIS beamline at the Swiss Light Source using the COPHEE spectrometer with two 40\,kV classical Mott detectors \cite{Hoesch2002spin} at a photon energy of 19\,eV.
Energy and angular resolutions were $\sim$60\,meV ($\sim$20\,meV in spin-integrated spectra) and 1.5$^{\circ}$ (0.5$^{\circ}$ in spin-integrated spectra), respectively.

\begin{figure} [ht]
	\includegraphics[width=\linewidth]{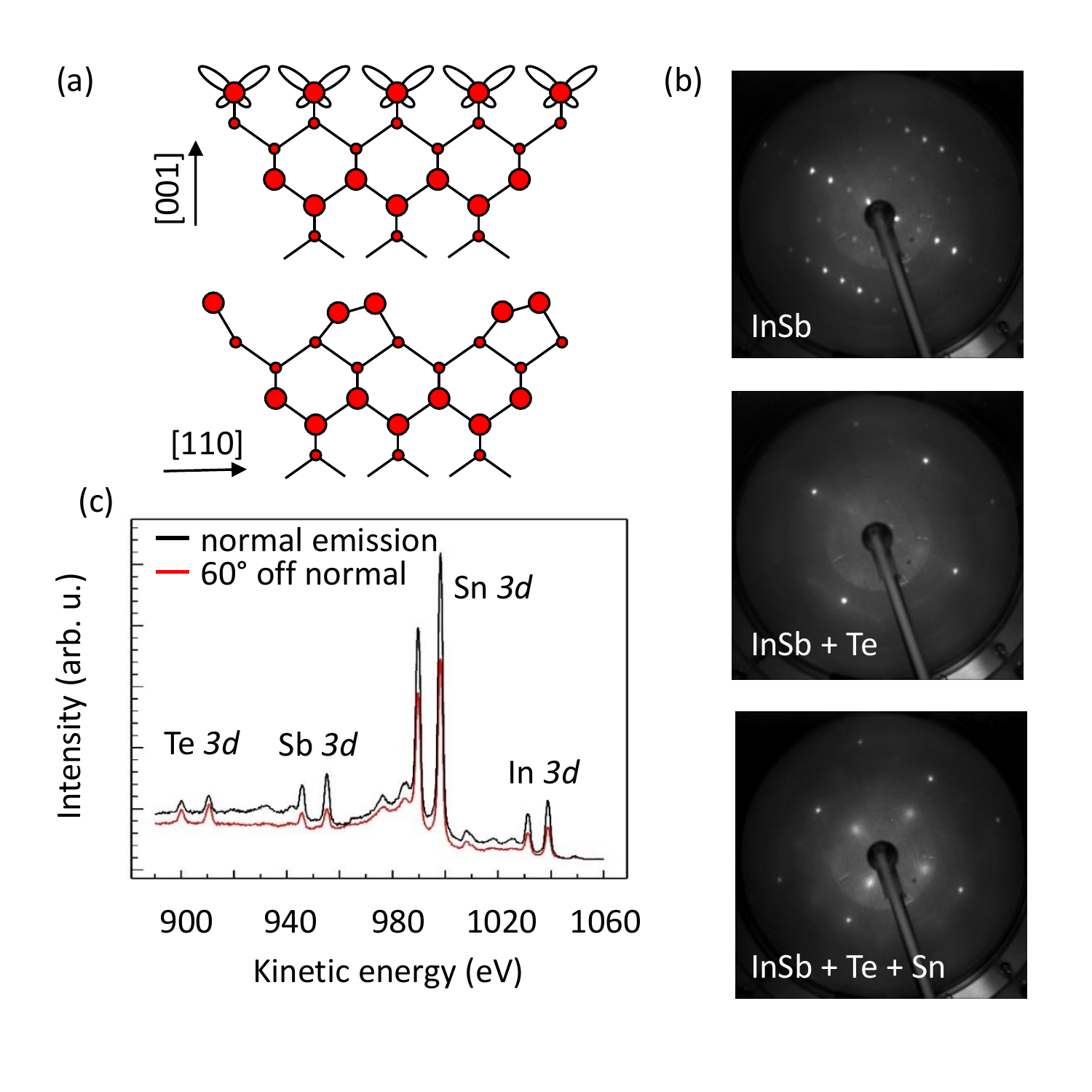}
	\caption{(a) Illustration of the surface-truncated diamond crystal structure (top) and the reconstructed surface (bottom). Smaller size of atoms denotes atoms situated further from the (110) plane along [1$\bar{1}$0] direction.
	(b) Evolution of the LEED pattern from a c(8$\times$2) surface reconstruction for \textit{in situ} prepared InSb (top), to a nearly (1$\times$1) for the Te buffer on InSb (center), to a (2$\times$1) double domain pattern for Sn grown on top of the buffer layer (bottom). 
	(c) Angle-dependent XPS (Al K$\alpha$) data.}
	\label{Fig1}
\end{figure}

\section{\label{Crystal}Atomic Surface Structure}
Since \aSn\ crystallizes in the diamond structure, we want to briefly review the structural basics of the (001) surface. If the bulk is truncated at the (001) surface, the topmost atoms miss their partners to saturate bonds. Being of $sp^3$-character, these dangling bonds are strongly directional and lie diagonally in the (110) plane. Local density approximation calculations show that the unreconstructed surface is energetically not favorable due to a very high surface energy of $1.530\, \mathrm{eV} / ($1$\times$1$)\ \mathrm{cell}$ \cite{Lu:1998ce}. According to Lu \etal\ \cite{Lu:1998ce}, a stabilization is achieved through the formation of asymmetric dimers, which leads to an energy gain of $0.618\, \mathrm{eV} / ($1$\times$1$)\ \mathrm{cell}$ [Fig.~\ref{Fig1}(a)]. This dimer formation is well known also for the Si(001) and Ge(001) surfaces \cite{Desjonqueres:1996th}, and it is the building block for many reconstructions observed in experiments.
For \aSn(001), the formation of reconstructions has been investigated by Yuen and coworkers, who found a dependence of the reconstruction type on the film thickness \cite{Yuen:1990hp}. In the regime of interest, i.e., below 200\,\AA, the formation of a (2$\times$1) reconstruction is favored. It shows up in a double-domain fashion since the dangling bonds of the truncated bulk are oriented in two orthogonal directions on neighboring surface terraces separated by a mono-atomic step. As can be seen in Fig.~\ref{Fig1}(b), the LEED images show an evolution from a c(8$\times$2)-reconstructed surface (top) for pristine InSb(001) after surface preparation to a (2$\times$1) reconstruction for the Sn-covered surface (bottom). We found that adding Te prior, during, or after the Sn deposition improves the crystal quality as it apparently acts as a surfactant. Depending on the amount of Te added, the dangling bonds of Sn become saturated and the formation of dimers is avoided, resulting in an unreconstructed (1$\times$1) surface structure \cite{Barfuss:2013by}. A similar effect is seen on the InSb surface after growth of a Te buffer layer [Fig.~\ref{Fig1}(b), center].
Figure~\ref{Fig1}(c) shows angle-dependent core-level spectra of a Sn film grown on InSb(001) where a Te buffer layer was added before Sn deposition.
The $3d$ states of all elements that contribute to the signal are situated next to each other in binding energy such that the qualitative dependence of the intensities on the emission angle can be easily followed.
The limited escape depth of photoelectrons leads to an enhanced surface sensitivity for off-normal emission where the escape depth is proportional to the cosine of the emission angle measured relative to the surface normal.
Naively, one would expect Te, Sb, and In to decrease in intensity for 60$^\circ$ off normal emission (light gray line) as Sn was deposited on top of the three.  As can be seen, this is not the case: with increasing surface sensitivity the intensity of Sb and In reduces, as expected. However, also the Sn intensity is decreasing, while the Te intensity appears unaltered, or slightly higher, for 60$^\circ$ off normal emission as compared to normal emission (0$^\circ$). This gives clear indication that Te segregates on top of the growing film, instead of buffering the interface between InSb and Sn.
Such behavior is typical of surfactants which alter the surface free energies \cite{Herman:1996te,Voigtlander:1995eo}.
We quantify the results from the angle-dependent XPS by calculating the relative spectral-weight change according to
\begin{eqnarray}
	\frac{A_{\rm 60}-A_{\rm NE}}{A_{\rm 60}+A_{\rm NE}},
	\end{eqnarray}
where $A_{\rm 60}$ and $A_{\rm NE}$ are the areas below each $3d_{5/2}$-peak for the measurement at 60$^\circ$ off normal emission and at normal emission, respectively. From each peak, a Shirley background has been removed prior to the analysis. The calculation gives an increase of $\sim$16\% for the Te state and a reduction of $\sim$19\% for the Sn state. Remarkably, the Sb state is reduced by $\sim$40\%, while the In state is reduced by less than $\sim$29\%. This indicates that diffusion of In into the film may persist despite the Te buffer layer.
Interdiffusion of In can be further enhanced by formation of metallic In islands that accompanies InSb substrate surface preparation via sputtering and annealing \cite{Farrow:1981il}. 
Furthermore, Te seems to act as a $n$-type dopant (as it does in group-IV semiconductors like Si and Ge \cite{Sze1981physics}), or, it is at least able to compensate intrinsic $p$-type doping that is likely to be caused by the interdiffusion of In atoms from the substrate into the film \cite{Barfuss:2013by}.

\begin{figure} [ht]
	\includegraphics[width=\linewidth]{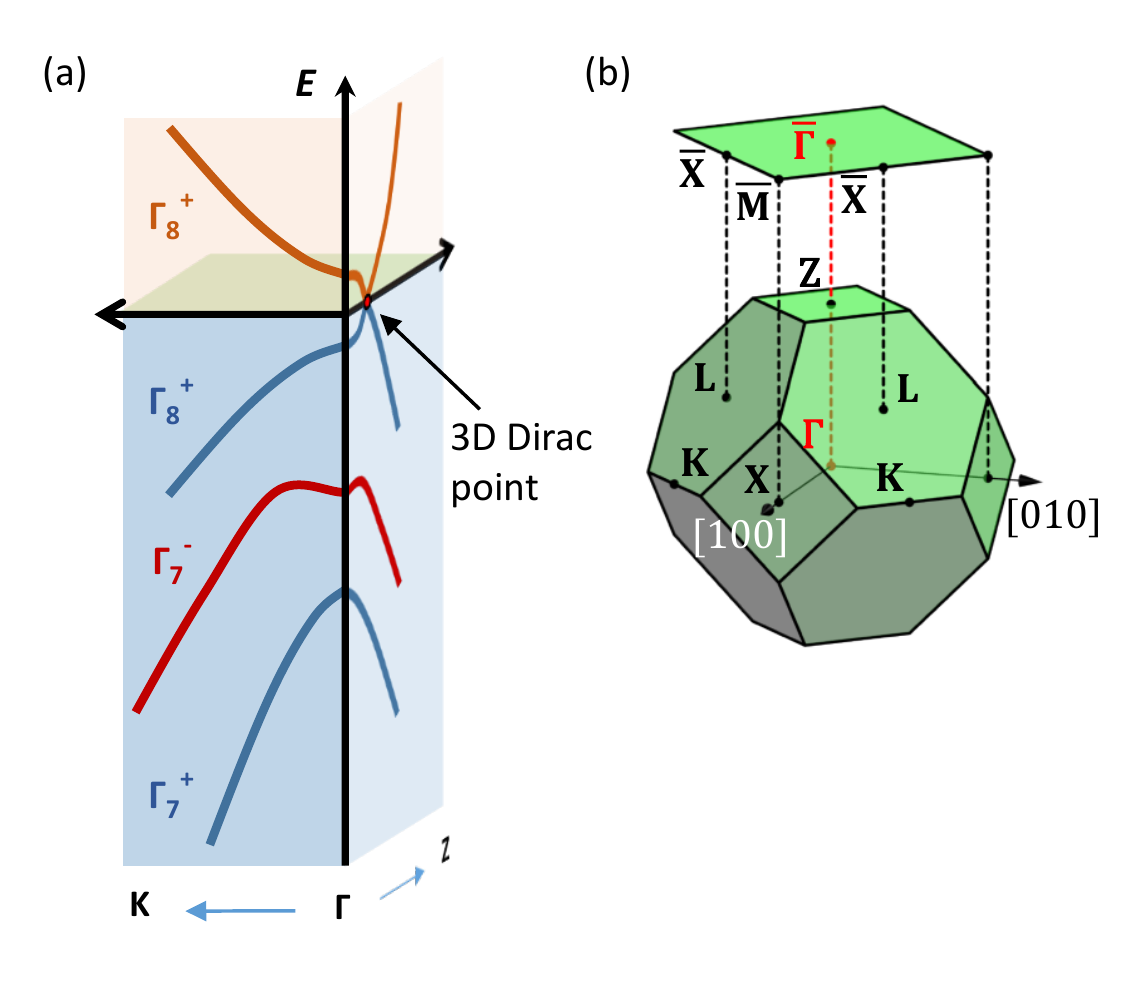}
	\caption{(a) Sketch of the bulk band structure of \aSn\ on InSb(001) along the in-plane $\Gamma$-K direction and the $\Gamma$-Z direction perpendicular to the (001) surface. (b) Bulk and surface Brillouin zone of strained \aSn\ with symmetry points labeled.}
	\label{BZ}
\end{figure}

\section{\label{Electronic}Electronic Structure}

\begin{figure*}  [!ht]
	\includegraphics[width=\linewidth]{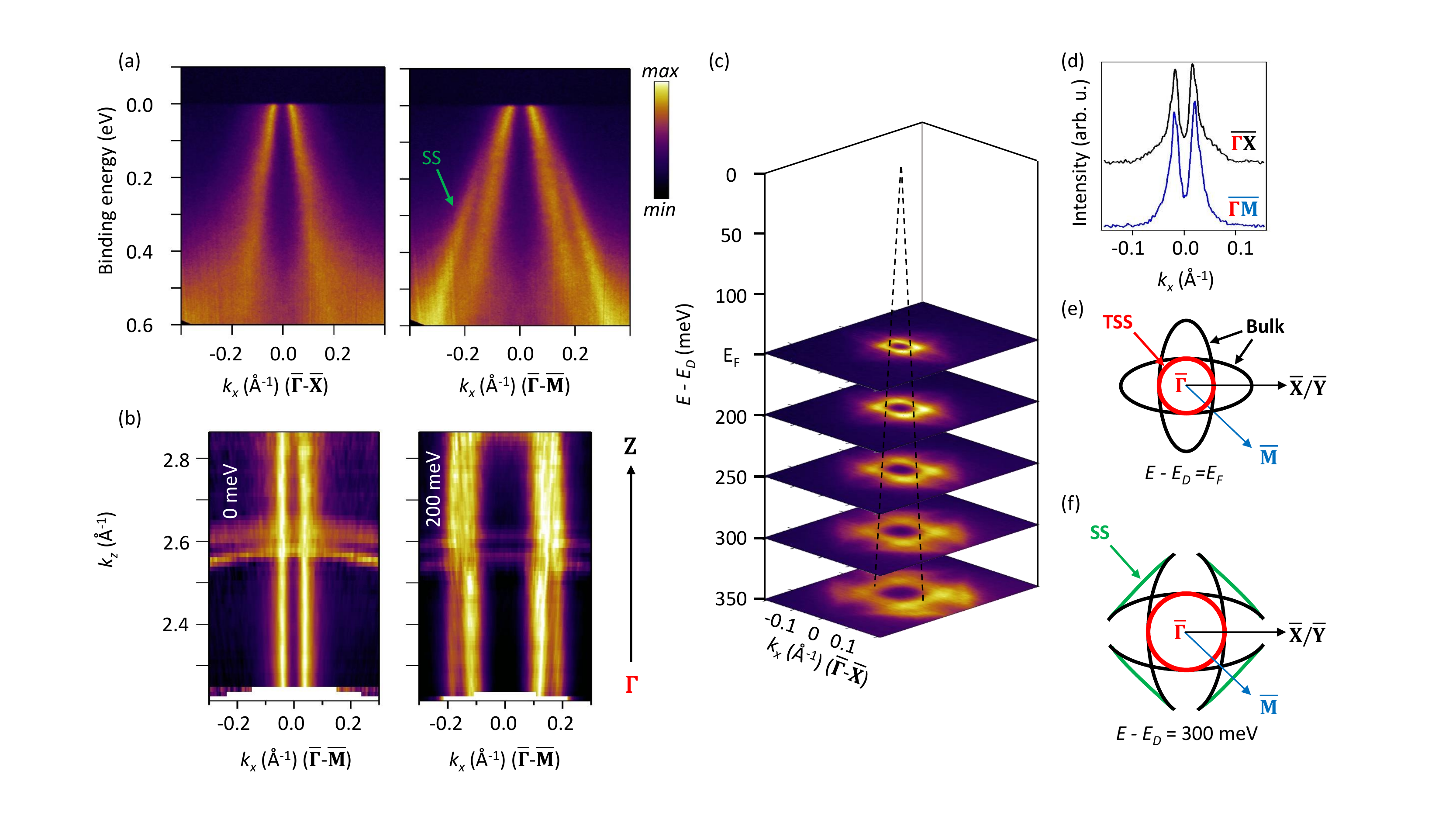}
	\caption{ Experimental electronic structure data.
		(a) Band maps of an as-grown \aSn\ film measured with $h \nu$=18\,eV at $T$=8\,K along the \GbarXbar\ direction (left) and \GbarMbar\ direction (right). (b) Photon energy scan between $h \nu$=18\,eV and $h \nu$=32\,eV taken along the the \GbarMbar\ direction at normal emission at \Ef\ (left) and  \Eb=250\,meV (right). Spectra have been normalized to have equal intensity for each \kz. The features that show up between \kz=2.5\,\invA\ and \kz=2.7\,\invA\ correspond to core levels that appear due to higher order excitations and can thus be ignored. (c) Stack of experimental CECs with black dotted lines following the TSS dispersion. (d) MDCs extracted at \Ef\ for the \GbarXbar\ and \GbarMbar\ directions. (e, f) Sketches of the experimental CECs for  different binding energies show the overlap of circular TSS (red) and elliptic bulk bands (black), as well as surface state SS (green).}
	\label{ARPES_standard}
\end{figure*}

Recently it was shown that the compressive strain in \aSn\ films induced by the InSb substrate opens only a local band gap at the $\Gamma$-point of the bulk Brillouin zone [BBZ, Fig.~\ref{BZ}(b)], which closes along the direction perpendicular to the (001) surface, i.e., along the $\Gamma$-Z direction [Fig.~\ref{BZ}(a)] \cite{Xu2017,Rogalev:2016tz}.
This gives rise to two three-dimensional Dirac points along the line Z-$\Gamma$-Z and defines \aSn\ as a topological Dirac semimetal.
We note that this gap closing does not alter the bulk topological properties as the strain induced band crossing occurs between the two \GEP bands that form the topmost valence and conduction band in unstrained \aSn. As both share the same parity eigenvalues, the local gap opening induced by strain does not change the \Z-invariant.
Note, that the size of the gap in the sketch of Fig.~\ref{BZ}(a) is exaggerated, and due to its small real size ($\sim$ 30\,meV) and photoemission \kperp\ broadening \cite{Strocov2003} we have no experimental resolution to clearly show the 3D Dirac points in ARPES.

We approach the topological transition by reviewing the electronic structure of InSb in comparison with \aSn.
Based on nonlocal pseudopotential calculations, Chelikowsky and Cohen were able to show that in InSb the \GSM band lies above \Ef\ for all time reversal symmetric momenta \cite{Chelikowsky:1976dv}. In contrast, the \GSM band is pushed below \Ef\ in \aSn \cite{Groves1963} due to mainly scalar relativistic effects that affect the s-electrons \cite{Cardona2010,Zhu2012}, thus giving rise to a change of the \Z-invariant \cite{Fu:2007ei}.
There is yet another band inversion in \aSn\ as the \GSP\ band is pushed below the \GSM\ band [Fig.~\ref{BZ}(b)] \cite{Rogalev:2016tz}. While this inversion has no effect on the \Z-invariant, the presence of a second TSS in \aSn\ was recently discovered, that indeed connects the \GSM and \GSP bands in agreement with this inversion \cite{Rogalev:2016tz}.

In Fig.~\ref{ARPES_standard}(a) we present energy dispersive maps of \aSn\ acquired at $T$=8\,K with an excitation energy of $h \nu$=18\,eV translating into a surface perpendicular momentum \kz = $2.33\times(2\pi/c$), assuming an inner potential of $V_0$=5.8\,eV (Ref.\ [\onlinecite{Rogalev:2016tz}]). The left-hand panel shows the dispersion along the \GbarXbar\ direction (note that due to the dual domain nature of the surface reconstruction in our samples the \Xbar\  and \Ybar\ points of the surface Brillouin zone (SBZ) can not be distinguished). At 0\,eV binding energy we observe a single sharp feature that crosses \Ef, corresponding to the TSS reported recently \cite{Barfuss:2013by,Ohtsubo:2013je}.
In addition, we observe a weak background intensity caused by the projected bulk \GEP band that also crosses \Ef, revealing a metallic behavior.
The apparent $p$-type doping is likely caused by diffusion of In from the substrate into the film \cite{Barfuss:2013by}, whereas the amount of Te in the buffer layer was not enough to compensate for it. From the momentum distribution curves (MDCs) taken at \Ef\ we extract the Fermi wavevector \kf\ for the \GbarXbar\ direction to $\sim$0.035\,\invA [Fig.~\ref{ARPES_standard}(d)]. 

When following the TSS dispersion to higher binding energy, a reduction in intensity, accompanied by a broadening, is observed. This broadening is ascribed to the presence of the bulk \GSM band, and a hybridization with it.
The hybridization is even more pronounced for the \GbarMbar\ direction as shown on the right-hand side of Fig.~\ref{ARPES_standard}(a). At \Ef\ we get a similar picture as along the \GbarXbar\ direction with the only intense feature being the TSS. For higher binding energies, a second feature is observed, which we will refer to as the conventional surface state (SS), and which gains intensity and sharpness with increasing binding energy.
In Fig.~\ref{ARPES_standard}(b) we show a photon energy scan between $h \nu$=18\,eV and $h \nu$=32\,eV, i.e., along the $\Gamma$-Z direction of the BBZ, while the surface is oriented with the \GbarMbar\ direction along the angle-dispersive axis of the analyzer. 

At the Fermi level [Fig.~\ref{ARPES_standard}(b), left] we observe a single intense feature that corresponds to the TSS, in agreement with the results of Fig.~\ref{ARPES_standard}(a). This feature appears as a straight line along the \kperp\ direction, which is clear proof of its two-dimensional character, i.e., its surface localization. At 250\,meV binding energy [Fig.~\ref{ARPES_standard}(b), right], in addition to the TSS, one can observe the second sharp feature that originates from the SS band [see Fig.~\ref{ARPES_standard}(a)]. Clearly, this additional feature shows no dispersion in $k_z$ and dresses the TSS with a constant separation in \kpara. We can therefore assign a two-dimensional character to this state as well.

To elucidate the topography of the states in more detail, we present a set of experimental CECs in Fig.~\ref{ARPES_standard}(c). Sketches of CECs are shown in Fig.~\ref{ARPES_standard}(e),(f). The energy-dependence of the intense TSS in the stack of CECs in Fig.~\ref{ARPES_standard}(c) is illustrated by the black guide lines, which suggest a Dirac-point energy of \Ed $\approx$150\,meV above \Ef. The shape of the Fermi contour can be resolved as a cloverleaf-like feature, resulting from two orthogonal elliptic contours that cross each other symmetrical about $\Gamma$. They originate from the bulk states and appear to be degenerate with the central TSS.
Since the intensities of the elliptic bulk features add up at the crossing points along the four \GbarMbar\ directions, the overall shape of the TSS appears to be quadratic. These crossing points are seen better at higher binding energies ($>$200\,meV) where the TSS shows a weaker intensity.

The quadratic shape would fulfill a strong nesting condition, with the possibility of an emerging spin-density wave \cite{Fu:2009ey}, however, the \kf\ values for TSS extracted from MDCs shown in Fig.~\ref{ARPES_standard}(d) are 0.035\,\invA and 0.037\,\invA\ for the \GbarXbar\ and \GbarMbar\ direction, respectively. The latter is far off the expected value of $\sqrt{2} k_F= 0.049$\,\invA\ along the \GbarMbar\ direction for a quadratic shape. Therefore, for the available binding-energy range, we establish a circular shape of the TSS-related Fermi contour, while it only appears to be quadratic due to overlap with bulk elliptic features.

\section{\label{DopStud}Surface doping study}

Previously, we have studied the effect of Te co-deposition during the growth of Sn on InSb(001) \cite{Barfuss:2013by}. It was shown that an enhanced flux of Te leads to a reduction of the intrinsic $p$-type doping, however, it has not been clarified whether Te prevents the outdiffusion of In atoms into the film, thus reducing the $p$-type doping, or whether Te itself acts as an electron donor. As already described above, samples in this study were grown with a Te buffer on InSb(001) instead, without an additional Te supply during the Sn deposition.


\begin{figure*}  [!htb]
	\includegraphics[width=\linewidth]{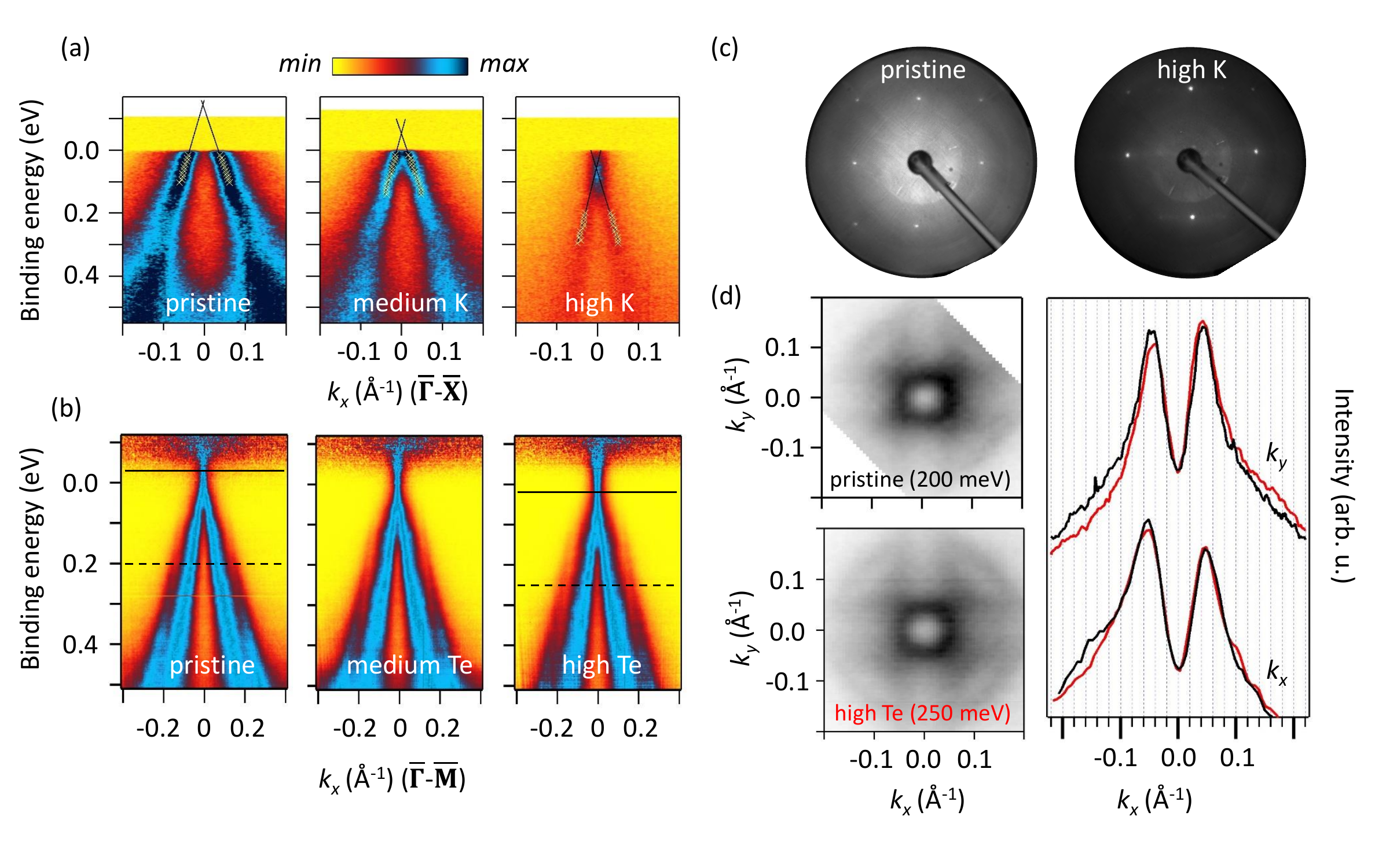}
	\caption{Study of K and Te adatom deposition.
		(a) ARPES band maps for the pristine sample (left), an intermediate K coverage (center, 3\,min. K deposition), and the final K coverage (right, 15\,min. K deposition) along the \GbarXbar\ direction with $h \nu$=18\,eV at 8\,K.
		(b) ARPES band maps for the pristine sample (left), an intermediate Te coverage (center), and high Te coverage (right) along the \GbarMbar\ direction at room temperature. Plots have been normalized to equal MDC intensity. 
		(c) While the as-grown sample shows a (2$\times$1) double-domain diffraction pattern in LEED (left), we observe a (1$\times$1) pattern at the end of K deposition study (right).
		(d) By comparing the CECs at the same energy relative to \Ed\ for the pristine and highly Te-doped sample, we find no obvious difference (left-hand panel). The right-hand panel compares MDCs for the two orthogonal directions $ k_x $ and $ k_y $ for pristine (\Eb=200\,meV, black) and high Te-doped (\Eb=250\,meV, red) samples. 
	}
	\label{Doping}
\end{figure*}

\textbf{K deposition.}
To be able to show that the TSS discussed with respect to Fig.~\ref{ARPES_standard} is indeed the TSS reported in previous studies \cite{Barfuss:2013by,Ohtsubo:2013je,RojasSanchez:2016gr}, we have conducted a series of K depositions on the surface of \aSn\ at $T$=8\,K, monitoring the changes in the band structure after each step. Figure \ref{Doping}(a) shows band maps for the pristine sample (left), an intermediate K coverage (center, 3\,min K deposition), and the final K coverage (right, 15\,min K deposition). To determine the TSS peak positions, MDCs were fitted with Lorentzian peaks in a small energy window. Subsequently, by fitting a straight line to the peak positions, one can extrapolate \Ed\ to quantify the shift of \Ef. For the pristine sample \Ed\ is found $\sim$150\,meV above \Ef.
One can clearly observe a shift of the Fermi level towards the Dirac point for the intermediate K dose. The band shift appears to be rigid since the group velocities determined from the linear approximation agree within the errors of our analysis. The Dirac point is now found $\sim$50\,meV above \Ef.
For the highest K dose, shown in the right-hand panel, the intensity from the Sn states is suppressed due to the highly disordered surface.
However, the extrapolation of the Dirac point matches well with the \textit{observed} Dirac point that shows an enhanced intensity.
The crossing point is found $\sim$65\,meV below \Ef, therefore we can quantify the total shift of \Ef\ induced by K to $\sim$215\,meV.
In addition to this shift, we observe a transition from a (2$\times$1) reconstruction of the pristine sample to an unreconstructed (1$\times$1) surface after the final K deposition in LEED [Fig.~\ref{Doping}(c)].

\textbf{Te deposition.}
The same transition was observed in LEED for Te surface deposition that has been conducted at room temperature (RT) in order to reduce possible cross-contamination due to otherwise extreme temperature changes between deposition and measurement steps.
The overall spectral quality becomes reduced at RT due to some k-space broadening of the states. On the other hand, due to the larger width of the Fermi-Dirac edge (of order $\sim$\ 4$ kT $) we are now able to probe also states slightly above the Fermi level. This spectral information can be enhanced through a normalization of the individual MDC intensities, as shown in Fig.~\ref{Doping}(b). For this sample, the initial \Ed\ is resolved at $\sim$15\,meV above \Ef, marked by the black line in the left panel. The reduction of the intrinsic $p$-type doping is ascribed to a higher initial Te amount in the buffer layer.
Two subsequent surface depositions of Te after the growth are shown in the middle and right panel of Fig.~\ref{Doping}(b). Again, we observe a clear shift of \Ef, and the total shift amounts to $\sim$50\,meV with \Ed=35\,meV for the highest Te content. We attribute this shift to an electron transfer from Te to Sn, providing evidence that the $n$-type doping reported in Ref.\ [\onlinecite{Barfuss:2013by}] is not only caused by a suppressed $p$-type doping from In diffusion.

Interestingly, the overall quality of the measurement appears slightly enhanced for the Te-doped sample. Especially the trivial surface state SS in the \GbarMbar\ direction [Fig.~\ref{ARPES_standard}(a)] is more pronounced. This is in stark contrast to the effect of K doping, where the overall quality was diminished. On the one hand, one may argue that Te forms an ordered overlayer that may smooth out the surface roughness of as-grown \aSn, in agreement with the surfactant character of Te. Potassium, on the other hand, may be unordered or form clusters, and an increased surface roughness could explain the low signal-to-noise ratio.

Figure~\ref{Doping}(d) shows CECs extracted at 200\,meV (top) binding energy in the pristine sample, and at 250\,meV (bottom), for the highly Te doped sample as marked by the white dashed lines in Fig.~\ref{Doping}(b). Hence, the energetic distance to \Ed\ is the same and allows for a comparison. As can be seen, the CECs are, despite some intensity differences, identical. 
In the right panel of Fig.~\ref{Doping}(d) we show MDCs extracted at $k_x$=0 (bottom) and $k_y$=0 (top) from the CECs to the left. Clearly, the peak maxima that stem from the TSS agree for both directions and doping levels. This underpins the argument that the surface reconstruction does not affect the TSS --- even though it is indeed strongly localized in the topmost layers \cite{Rogalev:2016tz}. This is in line with the findings by Ohtsubo \etal\ who showed that \aSn\ films covered with a (2$\times$1) reconstructed Bi layer exhibit no (2$\times$1) surface periodicity in the electronic structure \cite{Ohtsubo:2013je}.


\textbf{Fe deposition.}
Both Te and K overlayers have no magnetic moment and, therefore, the TRS of the system is preserved. As theory predicts, such perturbations to a TI are neither able to change the metallic behavior of the surface nor to destroy the TSS \cite{Hasan:2010ku,Qi:2011hb}, as long as the TSS bridges the gap between two different bulk bands. In contrast, the presence of a magnetic field in a topologically non-trivial system has been proposed to break the TRS and introduce a band gap in the TSS at \Ed. For two-dimensional surface states in 3D TIs many studies address this symmetry-breaking by deposition of ferromagnetic impurities or interfacing the TIs with a ferromagnetic overlayer \cite{Liu:2009ig,Gao:2009da,Wray:2010te,Hasan:2010ku,Garate:2010fh,Biswas:2010hp,Yu:2010fr,Mondal:2010kx,Chen:2010jv,Kong:2011ks,Valla:2012jz,Shelford:2012,Ye:2012df,Honolka:2012kp,Scholz:2012kz,Xu:2012fj,Watson:2013,Scholz:2013gl,Figueroa:2014,Baker:2015prox}.

\begin{figure} [ht]
	\includegraphics[width=\linewidth]{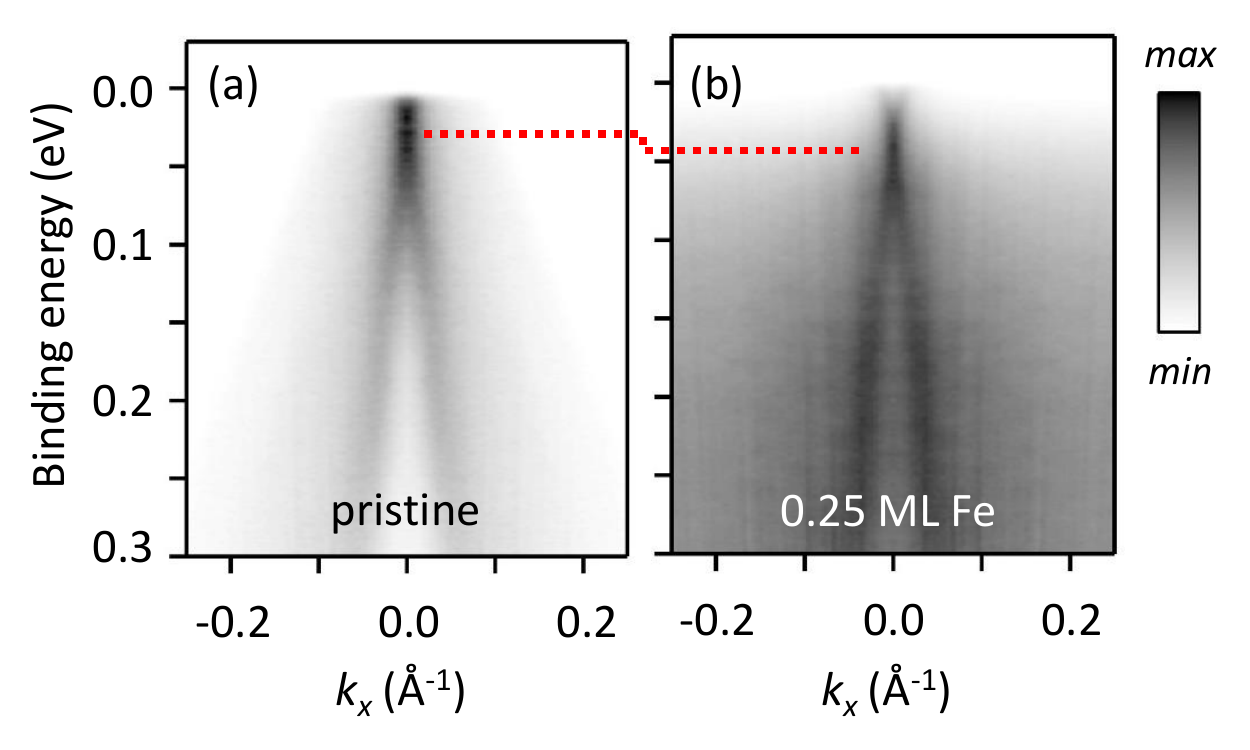}
	\caption{Study of Fe deposition. Pristine sample (left) and after deposition of $\sim$0.25 ML Fe (right).}
	\label{Doping_Fe}
\end{figure}

Figure~\ref{Doping_Fe} presents the results of Fe deposition on the surface of an \aSn\ film. The left panel shows the dispersion along \GbarXbar\ of an as-grown pristine \aSn\ sample, measured with photons of $h \nu$\,=\,18\,eV at $T$\,=\,8\,K. The TSS is clearly observed on top of the bulk background, as described in the previous section. The thickness of the Te buffer between substrate and \aSn\ film was adjusted such that the Dirac point is slightly, but clearly, below the Fermi level. 

In the right panel of Fig.~\ref{Doping_Fe}, we show the ARPES data on the same sample and measurement setup after deposition of 0.25\, monolayer (ML) Fe and subsequent annealing at $T$\,=\,373\,K. The bulk bands are not visible anymore due to the increased background signal from Fe \textit{d}-states. In contrast, the TSS appears very pronounced on top of this diffuse background signal. As outlined by the dashed line between the two panels of the figure, we again observe a clear shift of the Fermi level of the $n$-type sample. We quantify this shift by the position of the Dirac point prior and after Fe deposition to be $\sim$15\,meV. At \Ef, we clearly observe the two branches of the TSS at $\pm k$ that was not resolved in the pristine sample. Below \Ed, the TSS dispersion of the pristine and the Fe-covered sample agree well with each other, apart from the small energy shift. Most importantly, we find no sign of a band gap opening within our experimental resolution.

Interestingly, for a \Ed\ located close to the \Ef\, as in case of our pristine \aSn\ film, theory predicts that the TSS may mediate an alignment of the magnetic moments of the adatoms via a Ruderman-Kittel-Kasuya-Yosida (RKKY) interaction \cite{Liu:2009ig}. Typically, the RKKY interaction oscillates between the ferromagnetic and antiferromagnetic coupling as a function of 1/\kf, where \kf\ refers to the electronic state that mediates the coupling. Here, since \Ed\ was adjusted close to \Ef\ (\kf$\sim$0.005\invA), the period of the RKKY interaction oscillation $2\pi/$\kf\ increases to 100\, nm, which is much above the average distance between Fe adatoms. This means that the coupling between Fe adatoms should be ferromagnetic. However, we neither resolve an energy gap at \Ed\ nor do we observe a change in the band dispersion in terms of band mass or group velocity. Apart from the trivial absence of ferromagnetic order, the gap size could be still too small to be resolved with our given experimental resolution. Another reason for the gapless TSS dispersion could be intragap states induced by magnetic dopants \cite{sessi2016dual} that hide the clear gap opening in ARPES. In line with previous experimental findings in \BiSe\ \cite{Scholz:2012kz, Honolka:2012kp} and \BiTe\ \cite{Scholz:2013gl,Eelbo2014}, as well as theoretical considerations \cite{Biswas:2010hp}, we conclude that a simple 0.25\,ML Fe deposition onto the surface of \aSn\ is not able to open a resolvable band gap in the TSS.

\section{\label{SARPES}Spin-momentum locking}

\begin{figure*} [!]
	\includegraphics[width=\linewidth]{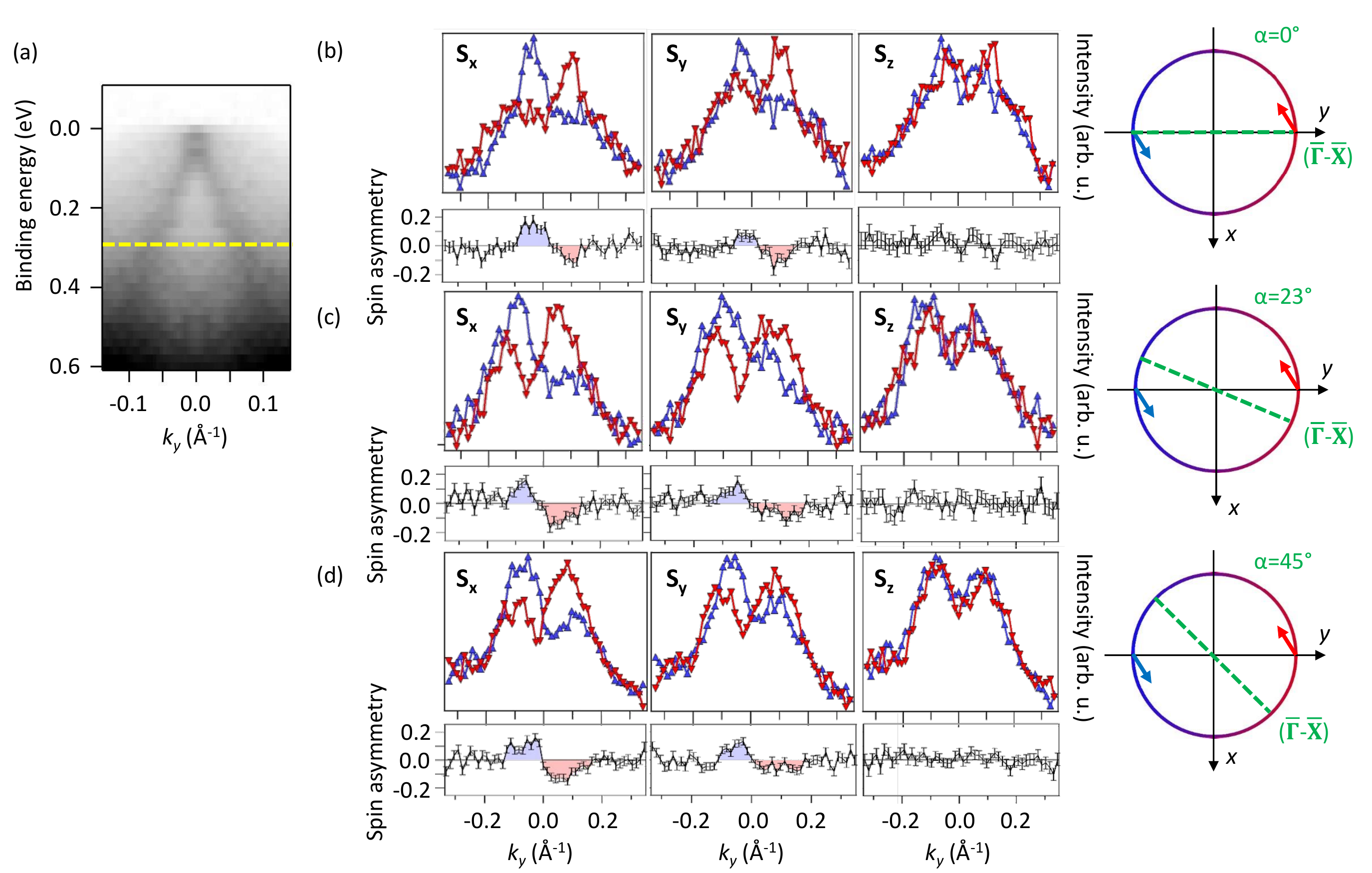}
	\caption{Spin-resolved MDCs for varying azimuthal angles $\alpha$. 
		(a) Spin-integrated ARPES map of an \aSn\ sample. The yellow dashed line shows the \Eb\ of spin-resolved MDCs in panels (b-d).
		(b-d) All three components (S$_x$, S$_y$, S$_z$) of the spin vector measured for three azimuthal rotations $\alpha$ = 0$^{\circ}$, 23$^{\circ}$ and 45$^{\circ}$, respectively. The circles represent CECs of TSS, while the red (blue) color corresponds to a positive (negative) sign of the S$_x$ component of the spin vector.
		}
	\label{Spin-ARPES}
\end{figure*}

Figure~\ref{Spin-ARPES} presents an in-depth study of the electron spin polarization of a sample that was optimized in terms of intrinsic doping, i.e., it has the Dirac point just slightly below the Fermi level as shown in Fig.~\ref{Spin-ARPES}(a). We have performed Mott-polarimetry of MDCs at \Eb$\sim$280\,meV as marked by the yellow dashed line in Fig.~\ref{Spin-ARPES}(a) and at different azimuthal rotations as marked by the green dashed lines on sketches on the right side of Fig.~\ref{Spin-ARPES}(b). Provided a 4-fold rotational symmetry, the selected cuts are representative for the whole surface state.

In Ref.\ [\onlinecite{Barfuss:2013by}] it was reported that a spin-vector was perpendicular to the momentum, as expected for an ideal Dirac cone. In Fig.\ \ref{Spin-ARPES}(b-d) we present all three vector components of the spin polarization (S$_x$, S$_y$, S$_z$)  measured in experimental geometries shown on the right side of each panel respectively. Note that the label $k_y$ refers to the experimental geometry which does not necessarily coincide with the sample high-symmetry directions. Since MDCs in Figs.\ \ref{Spin-ARPES}(b-d) were acquired by varying the tilt angle, we always move along the $k_y$-direction in $k$-space. In the same manner, the reference frame of the spin-polarization agrees always with the reference frame of the sample.
For all azimuthal rotations we get a similar result for all three components of the polarization. The polarization is most pronounced in the $x$-component and the reversal of the sign of polarization is clearly visible. $S_x$ is found to be positive for $-k_y$, and negative for $+k_y$, thus resulting in the opposite helicity as compared to TSS at \Ef\ in \BiSe \cite{Landolt2014spin}. 
The $z$-component shows vanishing polarization in agreement with the theoretical expectation that even rotational crystal symmetries confine the spin to the surface plane.

As a refinement of our previous result \cite{Barfuss:2013by}, we observe a finite polarization in the $y$-component as well. The effect is small, but significantly above the experimental error, and can not be simply explained with a small misalignment of the sample, i.e., an offset in $k_x$, since for that case the sign of the polarization should not reverse across $k_y$=0.
Therefore, our measurements qualitatively give a picture as shown in the sketches on the right side of Figs.\ \ref{Spin-ARPES}(b)-(d) where the arrows visualize the orientation of the spin (angular-momentum) vector. We see that the spin-momentum locking is not perpendicular, but is pointing inside of the constant energy contour with a finite component antiparallel to the momentum of the spin.
Resolving whether this is a pure spin effect or influenced by the coupling to the orbital momentum calls for further investigations, ideally assisted by one-step-photoemission theory.
We emphasize that despite the reduced $k$-resolution as compared to conventional ARPES, the spin-resolved MDCs are capable to distinguish states below the $k$-resolution as long as there is a difference in the polarization \cite{Dil2009spin}. In agreement with the results of the conventional ARPES [see Fig.~\ref{Doping}(d)], we do not observe a $\sqrt{2}$-factor between the MDC peak maxima at azimuth $\alpha$=45$^{\circ}$ [\GbarMbar\, Fig.~\ref{Spin-ARPES}(d)] and $\alpha$=0$^{\circ}$ [\GbarXbar\, Fig.~\ref{Spin-ARPES}(b)], giving further evidence that the constant energy shape of the TSS is close to circular.

\begin{figure}  [!h]
	\includegraphics[width=\linewidth]{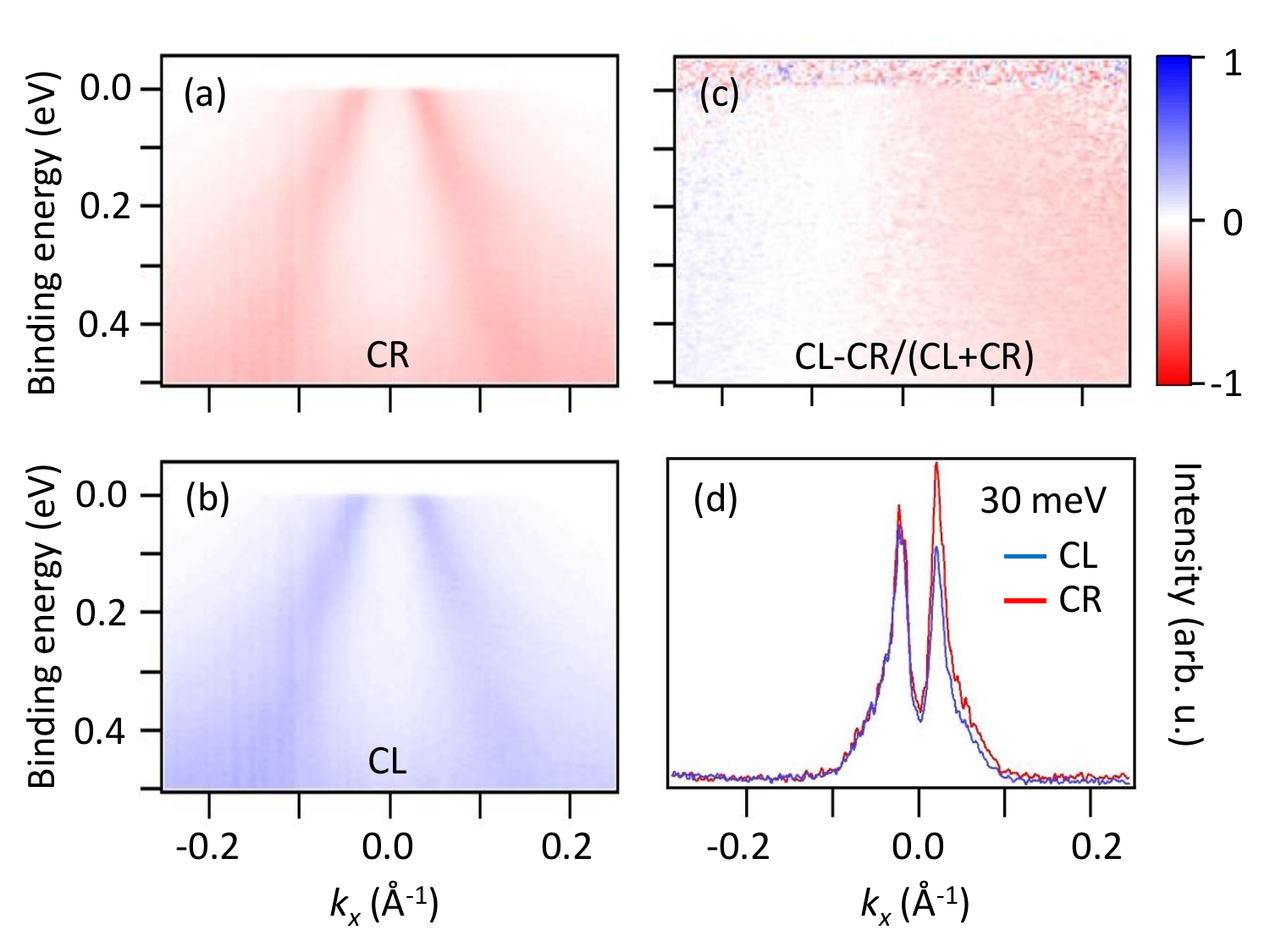}
	\caption{Analysis of the circular dichroism in \aSn\ at $h \nu$=18\,eV. Panels (a) and (b) show the intensity distribution measured with circular right- and left-polarized light, respectively. Panel (c) shows the normalized difference between (a) and (b). (d) Direct comparison of MDCs extracted at 30\,meV binding energy from (a) (red curve) and (b) (blue curve).}
	\label{Circ_Dich}
\end{figure}

To conclude the discussion of the spin-polarization, we show in Fig.~\ref{Circ_Dich} conventional ARPES measurements acquired with circularly polarized light. CDAD of photoelectrons has recently been suggested as a measure of spin in strong topological insulators \cite{Wang:2011fb,Park:2012fc}. In the study by Ohtsubo \etal\ \cite{Ohtsubo:2013je}, and also in the study by Liu \etal\ for HgTe \cite{Liu:2015ji}, this method has been applied and appears to confirm the measurements using Mott-polarimetry. As can be seen in our results presented in Fig.~\ref{Circ_Dich}, already a small difference in the excitation energy as compared to the results by Ohtsubo \etal, i.e., 18\,eV (here) vs. 19\,eV [\onlinecite{Ohtsubo:2013je}], leads to an almost perfect cancellation of the CDAD effect in \aSn.
We note that this is yet another confirmation that CDAD in spin-orbit coupled systems is no reliable stand-alone tool to measure the spin polarization of electronic states \cite{Scholz:2013fk,SanchezBarriga:2014fj}.

\section{\label{ImSigma}Many-body interactions}

Having confirmed the spin-momentum locking in the previous section, we now turn to a topic related to the spin-polarization of the surface state.
On the surface of a three-dimensional TI, 180$^{\circ}$ electron backscattering is forbidden as in case of the edge states in a two-dimensional TI. In contrast to the latter, scattering under angles smaller than 180$^{\circ}$ is only suppressed in three dimensions with a scatter probability that is dominated by the overlap of initial and final state spin states in the scattering process. According to theory \cite{Hatch:2011cx}, the probability of a hole introduced at a certain \kpara\ being filled with an electron of distinct \kpara\ continuously decreases between a scattering angle of 0$^{\circ}$ and $\pm$180$^{\circ}$ on a circular constant-energy contour. This holds for the case of a pure Rashba interaction, where the spin is locked perpendicular to the momentum. For the case presented here, the spin-dependent scattering probability should not be altered by the deviation from the perpendicular locking, as we observe only a moderate modification.

A much more significant impact is expected from the peculiarity that the TSS in \aSn\ is not situated in a projected bulk band gap, but exists completely degenerate with the projected \GEP state. One would naively expect strong hybridization effects as it is typical of such surface resonances. In this case, a hole created in the TSS by the photoemission process is readily filled with electrons from the bulk band. Such \textit{interband} scattering would not violate any conservation laws since the bulk states are not expected to be spin-polarized.
This would lead to a strong broadening of the spectral function of the TSS, amounting to very short lifetimes of the excited state.

\begin{figure} [!hb]
	\includegraphics[width=\linewidth]{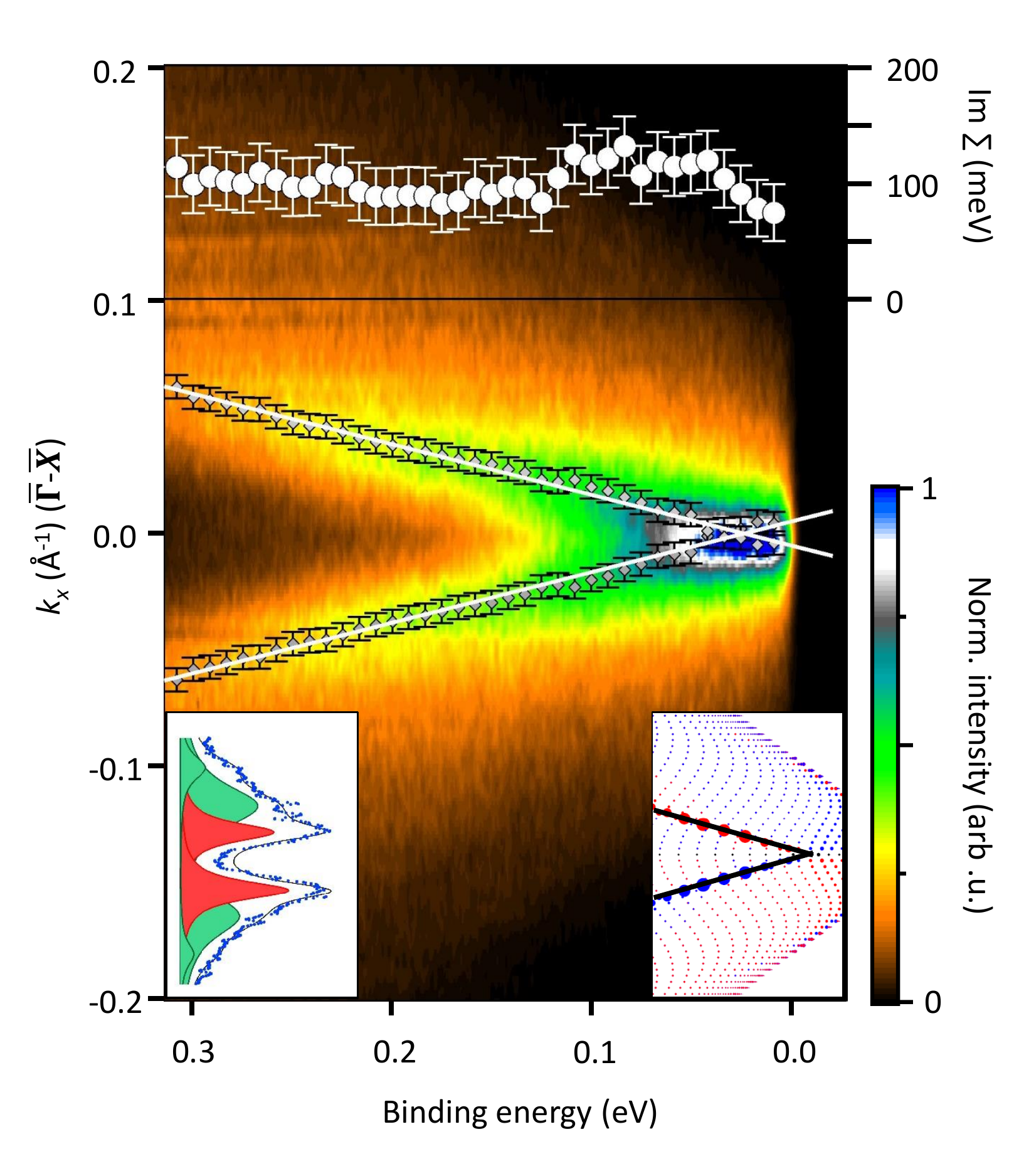}
	\caption{Analysis of MDCs of an \aSn\ sample. The straight white line is a linear fit to the peak positions (gray diamonds) of the TSS. The bottom-left inset shows an MDC at \Eb=0.31\,eV  fitted with 2 pairs of background peaks (green) and 1 pair of TSS peaks (red).
	The bottom-right inset shows a to-scale comparison of the linear fit to the experimental peak position (black line) with our $GW$-calculations. The size of the symbol is a measure of the surface localization, while the color represents the spin polarization (adapted from Ref.~[\onlinecite{Barfuss:2013by}]). The imaginary part of the electron self-energy (Im$\Sigma$) (white circles), extracted from the data, is shown in the top panel.
	}
	\label{Many_Body}
\end{figure}

Surprisingly, and as evident from the already discussed data, this is not the case in \aSn. The TSS appears pronounced and sharp against the bulk background of the \GEP band.
To quantify the subjective impression from the false-color plots, we analyze the peak width from MDCs through a fit of Lorentzian curves convolved with a fixed Gaussian width to account for the experimental resolution (0.005\,\invA). The results are shown in Fig.~\ref{Many_Body}. The analysis was applied to the data from a sample with increased Te amount in the buffer layer, which resulted in a Dirac point $\sim$35\,meV below the Fermi level. The bulk background was modeled with two pairs of peaks for the binding energy range above $\sim$90\,meV and with one pair of peaks below. Another pair of peaks has been fitted to the TSS in the whole binding energy range. Within each pair the peaks have been restricted to be symmetric about zero momentum and to have the same Lorentzian width at $\pm k$. These constraints gave meaningful results over almost the complete binding energy range. Only close to \Ed, where the routine was unable to fit two peaks to the TSS, we manually fixed the peak positions to get conclusive results. An example of MDC fitting is shown at the bottom-left inset in Fig.~\ref{Many_Body}.

In Fig.~\ref{Many_Body} the gray filled diamonds represent the peak positions of the TSS from the described routine and the black bars give the full width at half maximum (FWHM). For clarity reasons we show only an average over two neighboring points of the fit results. The raw data has been preprocessed such that the energy step agrees with the experimental resolution of $\sim$10\,meV.
In addition, we applied a linear regression to the peak positions as represented by the straight white line in the figure. As can be seen, the line fits very well to the fitted peak positions, confirming the linear dispersion of the TSS. We extract a group velocity of (4.5 $\pm$ 0.5)\,eV\AA, i.e., (6.8  $\pm$ 0.8) $\times$ 10$^5$\,m/s, which compares well with the Fermi velocity reported by Ohtsubo \etal\ of 7.3 $\times$ 10$^5$\,m/s \cite{Ohtsubo:2013je}. In the inset (bottom-right of Fig.~\ref{Many_Body}) we compare the group velocity from our analysis (black lines) with the one from theoretical \textit{GW}-calculations reported in Ref.\ [\onlinecite{Barfuss:2013by}], which are in excellent agreement.

To provide some insight into the decay mechanisms of created photoholes, we calculate the imaginary part of the electron self-energy (Im$\Sigma$) by multiplying the half width at half maximum (HWHM) of the Lorentzian peaks with the experimental group velocity \cite{Valla:1999ug}. The result is shown in the top panel of Fig.~\ref{Many_Body}.
At \Ef, we observe an offset of 80\,meV that is explained by the presence of impurities on the surface, which lead to an energy-independent inelastic scattering. We then observe an increase of \IMS to $\sim$120\,meV from \Ef\ to $\sim$50\,meV binding energy. A possible decay of photoholes in this range is through electron-hole pair creation or electron-phonon interaction. However, a characteristic quasi-particle kink is not clearly resolvable in the dispersion due to the overlap of the two broadened peaks from the TSS branches at $\pm k$.
Between binding energies of 50\,meV and 100\,meV, \IMS stays constant before a slight reduction to \IMS$\approx$100\,meV is observed. We note that this reduction appears in the vicinity of \Ed, and a detailed analysis of this behavior would be speculative due to the intricacies in the modeling. In this range we have reduced the number of peaks to model the bulk background from four to two. However, a similar decrease was observed for graphene on SiC and explained by the limited number of decays with enough momentum transfer for the creation of electron-hole pairs \cite{Bostwick:2007hn}.
At $\sim$150\,meV binding energy we identify a very small increase that would again fit to a decay by electron-hole pair creation.

So far, we have not taken into account the presence of the projected bulk states for the decay of photoholes. As mentioned above, one could expect a huge impact on the decay from these states. However, if we compare the average lifetimes of photoholes created in the TSS of \aSn\ ($\sim$3.3\,fs at \IMS$\approx$100\,meV) to those observed in other systems, where the surface state exists in a projected bulk band gap, like \BiSe\ ($\sim$11-30\,fs) \cite{Valla:2012jz} or \BiTe\ ($\sim$8-16\,fs) \cite{SanchezBarriga:2014it}, the difference is not so significant. Even in two-dimensional graphene on SiC, the lifetime of $\sim$5\,fs \cite{Bostwick:2007hn} is comparable to \aSn. In a recent study, we were able to show that the TSS we investigated is derived mainly from $p_z$-orbitals, while the \GEP band, the projection of which is degenerate with the TSS, has mainly $p_x+p_y$ character \cite{Rogalev:2016tz}. One may, therefore, argue about `orbital protection' against interband scattering for the TSS in \aSn. Moreover, this difference in the orbital character avoids hybridization and leads to a strong surface localization of the TSS despite its surface resonance character. In a real-space picture, one may argue that the density of states (DOS) at the surface originates mainly from the TSS since the bulk bands show a strong decay into the vacuum. Hence, the TSS DOS alone is left as a phase space for scattering events, which might explain the surprisingly high lifetimes of the photoholes.

\section{\label{Conclude}Conclusions}
In conclusion, we have analyzed the electronic structure of the TSS in strained \aSn\ grown on InSb(001) by spin- and angle-resolved photoemission. 
We found that the TSS has an almost ideal Dirac cone shape with a circular Fermi contour.
With the help of photon energy scans, we identified a second two-dimensional feature that appears only away from the \GbarXbar\ direction in $k$-space.
We have shown that Te, as well as K, act as electron donors if deposited on the surface of \aSn\, and they allow to shift the Dirac point by at least $\sim$50\,meV and $\sim$215\,meV, respectively. Most importantly, we have demonstrated that deposition of 0.25\,ML of ferromagnetic Fe induces a slight $n$-type doping, however, no resolvable gap in ARPES.
Moreover, we analyzed the spin-momentum locking of TSS electrons with spin-resolved photoemission and showed that the spin is captured within the surface plane.
It has a finite component that is antiparallel to the momentum.
Although the TSS is clearly spin-polarized, we observed that the circular dichroism in the angular distribution vanishes.
Finally, we have shown that the quasi-particle lifetimes of photoholes created in the TSS of \aSn\ compare very well with those of other TIs --- despite the peculiarity of being fully degenerate with the bulk states. We argue that the long lifetimes are due to an `orbital protection' (low hybridization) against interband scattering between the TSS and background of bulk bands.

\section{\label{Aknowledge}Acknowledgments}
We thank A. Fleszar for support with calculations and discussions on the manuscript. This work was supported by the Deutsche Forschungsgemeinschaft (DFG) under Grant SCHA1510/5, the SPP 1666 Priority Program ”Topological Insulators“, the DFG Collaborative Research Center SFB 1170 ”ToCoTronics” in W\"urzburg. Diamond Light Source (Didcot, UK) is gratefully acknowledged for beamtime under proposals SI10289, SI10244, SI12892, and SI15285. LCM and LBD acknowledge financial support from EPSRC (UK), and LBD also from the Science and Technology Facilities Council (UK).

\bibliographystyle{apsrev4-1.bst}

\end{document}